\documentclass[a4paper,11pt]{article}
\usepackage{jcappub} 
\usepackage{lineno}
\usepackage[margin=5pt,caption=false]{subfig}
\usepackage{slantsc}
\usepackage{comment}
\usepackage{threeparttable}
\usepackage{booktabs}
\usepackage[figuresright]{rotating}
\usepackage{textcase}
\usepackage{multirow}
\newcommand{\be}{\begin{equation}}
\newcommand{\ee}{\end{equation}}
\newcommand{\bea}{\begin{eqnarray}}
\newcommand{\eea}{\end{eqnarray}}
\newcommand{\hunit}{$\rm{km \ s^{-1} \ Mpc^{-1}}$}
\newcommand{\lcdm}{$\Lambda$CDM}
\newcommand{\pcdm}{$\phi$CDM}
\usepackage{array}

\newcommand{\hiig}{H\,\textsc{ii}G}
\newcommand{\hii}{H\,\textsc{ii}}
\newcommand{\Om}{\Omega_{m0}}
\newcommand{\Ok}{\Omega_{k0}}
\newcommand{\om}{$\Omega_{m0}$}
\newcommand{\ok}{$\Omega_{k0}$}
\newcommand{\wx}{$w_{\rm X}$}
\newcommand{\wX}{w_{\rm X}}

\newcommand{\mii}{Mg\,\textsc{ii}}

\newcommand{\civ}{C\,\textsc{iv}}

\newcommand{\obh}{\Omega_{b}h^2}
\newcommand{\och}{\Omega_{c}h^2}
\newcommand{\onh}{\Omega_{\nu}h^2}
\newcommand{\obhs}{$\Omega_{b}h^2$}
\newcommand{\ochs}{$\Omega_{c}h^2$}

\usepackage[T1]{fontenc}
\usepackage{scalerel}
\usepackage{tikz}
\usetikzlibrary{svg.path}

\definecolor{orcidlogocol}{HTML}{A6CE39}
\tikzset{
  orcidlogo/.pic={
    \fill[orcidlogocol] svg{M256,128c0,70.7-57.3,128-128,128C57.3,256,0,198.7,0,128C0,57.3,57.3,0,128,0C198.7,0,256,57.3,256,128z};
    \fill[white] svg{M86.3,186.2H70.9V79.1h15.4v48.4V186.2z}
                 svg{M108.9,79.1h41.6c39.6,0,57,28.3,57,53.6c0,27.5-21.5,53.6-56.8,53.6h-41.8V79.1z M124.3,172.4h24.5c34.9,0,42.9-26.5,42.9-39.7c0-21.5-13.7-39.7-43.7-39.7h-23.7V172.4z}
                 svg{M88.7,56.8c0,5.5-4.5,10.1-10.1,10.1c-5.6,0-10.1-4.6-10.1-10.1c0-5.6,4.5-10.1,10.1-10.1C84.2,46.7,88.7,51.3,88.7,56.8z};
  }
}

\newcommand\orcidicon[1]{\href{https://orcid.org/#1}{\mbox{\scalerel*{
\begin{tikzpicture}[yscale=-1,transform shape]
\pic{orcidlogo};
\end{tikzpicture}
}{|}}}}
\usepackage{amsmath,bm}
\usepackage{amssymb}
\usepackage{fnpct}

\arxivnumber{2502.08429} 
\title{\boldmath Testing the consistency of new Amati-correlated gamma-ray burst dataset cosmological constraints with those from better-established cosmological data}

\author[a,b]{Shulei Cao$^{\orcidicon{0000-0003-2421-7071}}$,}
\author[a]{Bharat Ratra$^{\orcidicon{0000-0002-7307-0726}}$}
\affiliation[a]{Department of Physics, Kansas State University,\\
116 Cardwell Hall, Manhattan, KS 66506, USA}
\affiliation[b]{Department of Physics, Southern Methodist University,\\
3215 Daniel Ave, Fondren Science Building, Dallas, TX 75205, USA}

\emailAdd{shuleic@smu.edu, ratra@ksu.edu}

\abstract{Gamma-ray bursts (GRBs) are promising cosmological probes for exploring the Universe at intermediate redshifts ($z$). We analyze 151 Fermi-observed long GRBs (datasets A123 and A28) to simultaneously constrain the Amati correlation and cosmological parameters within six spatially flat and nonflat dark energy models. We find that these datasets are standardizable via a single Amati correlation, suggesting their potential for cosmological analyses. However, constraints on the current value of the nonrelativistic matter density parameter from A123 and the combined A123 + A28 data exhibit $>2\sigma$ tension with those derived from a joint analysis of better-established Hubble parameter [$H(z)$] and baryon acoustic oscillation (BAO) data for most considered cosmological models. This tension indicates that these GRB data are unsuitable for jointly constraining cosmological parameters with better-established $H(z)$ + BAO and similar data. Although the A28 data constraints are consistent with the $H(z)$ + BAO data constraints, its limited sample size (28 GRBs) and high intrinsic scatter ($\sim0.7$) diminishes its statistical power compared to existing datasets.}

\begin{document}
\maketitle
\flushbottom

\section{Introduction}

Compelling evidence from diverse astronomical observations indicates that the cosmological expansion is currently accelerating. The dominant explanation of this acceleration attributes it to a gravitational effect of dark energy, a hypothetical substance with negative pressure. The spatially flat \lcdm\ model \citep{peeb84}, widely adopted within the cosmology community, postulates that dark energy is a cosmological constant ($\Lambda$), accounting for approximately 70\% of the Universe's present energy density. However, recent studies have suggested potential tensions with this standard model (see, for example, \citep{PerivolaropoulosSkara2021,Morescoetal2022,Abdallaetal2022,Hu:2023jqc}), motivating investigations into alternative cosmological models that allow for dark energy dynamics or incorporate spatial curvature. In this work we explore some of these alternative models to assess the potential of gamma-ray bursts (GRBs) as standardized cosmological probes.

Better-established cosmological observations primarily probe the low-redshift Universe ($z < 2.3$; \citep{CaoRatra2023}) or the epoch of recombination at $z \sim 1100$ (as observed in the cosmic microwave background; \citep{planck2018b}). GRBs, detected at redshifts extending to $z\sim8.2$, are a potentially valuable tool for exploring the largely uncharted intermediate redshift regime. Through empirical correlations, certain classes of GRBs are potentially standardizable probes of cosmological expansion \citep{Dirirsa2019, KhadkaRatra2020c, Khadkaetal_2021b, Caoetal_2021, ZhaoXia2022, DainottiNielson2022, CaoDainottiRatra2022b, Liuetal2022b, GovindarajDesai2022, Jiaetal2022, Liangetal2022, Singhetal2024, Kumaretal2023, Lietal2023, Muetal2023, Lietal2023b, Xieetal2023, Zhangetal2023, Favaleetal2024, CaoRatra2024b, Lietal2024, Bargiachietal2025, Dengetal2025}. To mitigate the GRB circularity problem and to assess the dependence of these correlations on the assumed cosmological model---a crucial step towards standardization---we conduct simultaneous analyses of both correlation and cosmological parameters across a range of cosmological models \citep{KhadkaRatra2020c, Khadkaetal_2021b, Caoetal_2021}.

Earlier investigations show that a sample of 118 Amati-correlated GRBs (hereafter referred to as A118), distinguished by reduced intrinsic scatter, exhibit a correlation between GRB properties that is robust against variations in the cosmological background. In addition to the Amati relation, the 2D and 3D Dainotti relations have also been extensively studied \citep{CaoKhadkaRatra2022,CaoDainottiRatra2022,CaoDainottiRatra2022b}. While \cite{CaoDainottiRatra2022b} demonstrated that X-ray GRB data strongly favor the 3D Dainotti relation over the 2D one, the latter remains viable, as suggested by \cite{Tianetal2023}, when the slope of internal plateaus satisfies $\leq 0.5$. In \cite{CaoDainottiRatra2022}, we conducted joint analyses of 101 Amati-correlated GRBs (A101) and 50 Platinum GRBs. As the A123 dataset results in this study are inconsistent with those from better-established $H(z)$ + BAO data, and as the A28 dataset (compared to A118) contains very few GRBs, we do not perform joint analyses of the A123 or A28 data with other GRB data. This suggests the viability of standardizing the A118 sample for use in cosmological studies \citep{Khadkaetal_2021b, LuongoMuccino2021, CaoKhadkaRatra2021, CaoDainottiRatra2022, Liuetal2022, CaoRatra2022}. While the cosmological constraints derived from the A118 GRB dataset are consistent with those obtained from better-established methods, the precision of the A118 GRB constraints is currently much lower than those from better-established techniques. However, combining the A118 dataset with these more precise measurements can (presently only weakly) refine constraints on cosmological parameters. The limited size of the A118 sample is responsible for the weakness of the GRB constraints, and highlights the need for larger, standardizable GRB datasets to both improve the precision of cosmological measurements and to enhance our understanding of GRB characteristics. Forthcoming higher-quality GRB catalogs from missions such as the recently-launched SVOM \citep{Atteiaetal2023}, the Einstein Probe \citep{EinsteinProbe2015}, the Roman Space Telescope \citep{Spergeletal2015}, the Rubin Observatory Legacy Survey of Space and Time \citep{LSST2009}, ULTIMATE-SUBARU \citep{ULTIMATESUBARU2014}, and the proposed THESEUS mission \citep{Amatietal2021} are expected to deliver such improved datasets. To further enhance cosmological GRB datasets, employing redshift estimates from Refs.~\cite{Dainottietal2024a,Dainottietal2024b,Narendraetal2024}, along with the redshift classification from Ref.~\cite{Dainottietal2025}, can more than double the sample size. In addition, lightcurve reconstruction techniques, as detailed in Refs.~\cite{Dainottietal2023,Kaushaletal2024}, can improve lightcurve properties, leading to better data quality and enabling better cosmological constraints with a reduced number of GRBs.

Applying the same analysis methodology used for the A118 GRB sample, here we investigate a new, purely Fermi, dataset of 151 Amati-correlated long GRBs (123 GOLD + 28) recently compiled by Ref.~\cite{WangLiang2024}. This new sample, representing a 28\% increase in size compared to A118, allows us to assess the standardizability of a purely Fermi GRB compilation and to compare the resulting cosmological constraints with those derived from better-established cosmological probes, such as baryon acoustic oscillation (BAO) observations and Hubble parameter [$H(z)$] measurements.

The technique developed to study the standardizability of GRBs \citep{KhadkaRatra2020c, Khadkaetal_2021b, Caoetal_2021}, has also recently been used to study and develop or reject several other potential cosmological probes. Reverberation mapping of \mii\ and \civ\ quasars (QSOs) or active galactic nuclei, extending to redshifts of $\sim 3.4$, offers a promising avenue of research \citep{Czernyetal2021, Zajaceketal2021, Khadkaetal_2021a, Khadkaetal2021c, Khadka:2022ooh, Cao:2022pdv, Czerny:2022xfj, Caoetal2024, Caoetal2024b}. The standardization technique \citep{KhadkaRatra2020c, Khadkaetal_2021b, Caoetal_2021} has demonstrated the potential for these QSO measurements to be standardized. Higher-redshift \hii\ starburst galaxies (\hiig), observed up to $z\sim2.5$ \citep{G-M_2019, CaoRyanRatra2020, GM2021, CaoRyanRatra2021, CaoRyanRatra2022, Johnsonetal2022}, are another potential cosmological tool. However, application of our standardization method \citep{KhadkaRatra2020c, Khadkaetal_2021b,   Caoetal_2021} to the most-recent \hiig\ compilation \citep{GM2021} reveals a critical challenge: while both low- and high-redshift \hiig\ subsamples can be individually standardized, they follow distinct correlations, precluding joint standardization \citep{CaoRatra2024a}. This finding challenges prior analyses of these data \citep{CaoRatra2024a}. QSO observations of X-ray and UV fluxes, probing redshifts up to approximately 7.5, have also been explored as potential cosmological indicators based on similar correlation analyses \citep{RisalitiLusso2015, RisalitiLusso2019, KhadkaRatra2020a, Yangetal2020, KhadkaRatra2020b, Lussoetal2020, KhadkaRatra2021, KhadkaRatra2022, DainottiBardiacchi2022}. Analysis of the latest QSO flux catalog \citep{Lussoetal2020}, however, indicates that these QSOs cannot be reliably standardized for cosmological purposes, as their correlation exhibits both cosmological model dependence and redshift dependence \citep{KhadkaRatra2021, KhadkaRatra2022}. This lack of standardizability limits their utility in cosmological studies \citep{KhadkaRatra2021, KhadkaRatra2022, Petrosian:2022tlp, Khadka:2022aeg, Zajaceketal2024, Wangetal2024}.

Here we find that independent analyses of the 123 GRBs (A123) and 28 GRBs (A28) datasets from Ref.~\cite{WangLiang2024} yield consistent values for both Amati correlation and cosmological parameters across different cosmological models. This consistency establishes, for the first time (\cite{WangLiang2024} did not consider this point), their standardizability through a single, cosmology-independent Amati correlation and also justifies a joint analysis of the combined (A123 + A28) dataset. However, while the A28 constraints are consistent with the $H(z)$ + BAO constraints, the small sample size of A28 (only 28 GRBs) and its higher intrinsic scatter ($\sim0.7$) make it a less compelling dataset for joint analyses compared to the much larger A118 sample \citep{Khadkaetal_2021b} with an intrinsic scatter $\sim0.4$. Furthermore, the current value of nonrelativistic matter density parameter \om\ constraints from both the A123 and A123 + A28 datasets for the four flat and nonflat \lcdm\ and \pcdm\ models show $>2\sigma$ tension with the better-established $H(z)$ + BAO data constraints. This tension means that both A123 and A123 + A28 datasets are not suitable for cosmological analysis in conjunction with better-established probes, such as type Ia supernova (SNIa) data, as done in Ref.~\cite{WangLiang2024}. This is also the case for the J220 \citep{Jiaetal2022} GRB dataset, Ref.~\cite{CaoRatra2024b}, and reaffirms that the A118 sample \citep{Khadkaetal_2021b, LuongoMuccino2021, CaoKhadkaRatra2021, CaoDainottiRatra2022, Liuetal2022, CaoRatra2022} is the most suitable GRB dataset for cosmological purposes.

This paper is organized as follows. We begin with a brief overview of the cosmological models in Sec.~\ref{sec:model}, followed by a description of the datasets used in Sec.~\ref{sec:data}. Section \ref{sec:analysis} briefly introduces our analysis methodology, and Section \ref{sec:results} presents our key results. We conclude in Sec.~\ref{sec:conclusion}.

\section{Cosmological models}
\label{sec:model}

In previous studies (see Refs.\ \cite{Khadkaetal_2021b, LuongoMuccino2021, CaoKhadkaRatra2021, CaoDainottiRatra2022, Liuetal2022, CaoRatra2022, CaoRatra2024b} and references therein), GRB data have been analyzed under the Amati ($E_{\rm p}-E_{\rm iso}$) relation \cite{Amatietal2002}, which connects the GRB's rest-frame peak photon energy with the isotropic energy, calculated from the observed bolometric flux of the GRB (see Sec.~\ref{sec:data} below). This paper focuses on the GRB data compilation of Ref.\ \cite{WangLiang2024}, exploring it in the context of six distinct relativistic dark energy models, both spatially flat and nonflat.\footnote{Recent discussions on spatial curvature constraints can be found in Refs.\ \cite{Oobaetal2018b, ParkRatra2019b, DiValentinoetal2021a, ArjonaNesseris2021, Dhawanetal2021, Renzietal2022, Gengetal2022, MukherjeeBanerjee2022, Glanvilleetal2022, Wuetal2023, deCruzPerezetal2023, DahiyaJain2022, Stevensetal2023, Favaleetal2023, Qietal2023, deCruzPerez2024, ShimonRephaeli2024, WuZhang2024}.} We want to determine whether these GRB data align with the $E_{\rm p}-E_{\rm iso}$ correlation independent of the assumed cosmological model, thus potentially qualifying them as standardizable. Additionally, we derive cosmological parameter constraints using these GRB data, to assess consistency with cosmological parameter constraints derived from other datasets. For this analysis, the Hubble parameter $H(z)$ for each model is computed as a function of redshift $z$ and cosmological model parameters, based on the first Friedmann equation within the framework of general relativity and the Friedmann-Lema\^{i}tre-Robertson-Walker metric.

The cosmological models we use here assume one massive and two massless neutrinos, an effective relativistic species number $N_{\rm eff} = 3.046$, and a total neutrino mass $\sum m_{\nu}=0.06$ eV. This yields a present-day nonrelativistic neutrino physical energy density parameter $\onh=\sum m_{\nu}/(93.14\ \rm eV)$, where $h$ is the Hubble constant scaled by 100 \hunit. The present-day nonrelativistic matter density parameter $\Om$ is then given by $\Om = (\onh + \obh + \och)/{h^2}$, where \obhs\ and \ochs\ are the baryonic and cold dark matter physical energy density parameters today. Given our focus here on late-time data, we neglect photon contributions to the cosmological energy budget.

We utilize both \lcdm\ and XCDM models; the latter generalizes \lcdm\ by allowing dark energy density to vary over time (though not spatially), parameterized by a constant equation of state parameter $w_{\rm DE}=p_{\rm DE}/\rho_{\rm DE}$, where $p_{\rm DE}$ and $\rho_{\rm DE}$ are the dark energy pressure and density, respectively. For \lcdm\ $w_{\rm DE} = -1$, whereas XCDM allows different $w_{\rm DE}$ values, leading to a modified Friedmann equation
\be
\label{eq:HzLX}
H(z) = H_0\sqrt{\Om\left(1 + z\right)^3 + \Ok\left(1 + z\right)^2 + \Omega_{\rm DE0}\left(1+z\right)^{3(1+w_{\rm DE})}}.
\ee
Here \ok\ is the current spatial curvature density parameter and $\Omega_{\rm DE0} = 1 - \Om - \Ok$ is the dark energy density parameter today. In \lcdm\ $\Omega_{\rm DE0} =\Omega_{\Lambda}$, while in XCDM $\Omega_{\rm DE0} =\Omega_{\rm X0}$, due to the dynamical X-fluid with equation of state parameter $w_{\rm X}$. Given that GRB data do not constrain $H_0$ or $\Omega_{b}$, we set $H_0=70$ \hunit\ and $\Omega_{b}=0.05$ in the GRB only data analyses. \lcdm\ model parameters constrained in our GRB only data analyses are $\Om$ and $\Ok$, and XCDM parameters constrained are $\Om, \wX$, and $\Ok$. For $H(z)$ + BAO data, we constrain $H_0, \obh\!, \och\!$, and $\Ok$ in \lcdm\ with an additional $\wX$ in XCDM. In the spatially flat cases $\Ok=0$.

We also utilize \pcdm\ models \cite{peebrat88,ratpeeb88,pavlov13},\footnote{See Refs.\ \cite{ooba_etal_2018b, ooba_etal_2019, park_ratra_2018, park_ratra_2019b, park_ratra_2020, Singhetal2019, UrenaLopezRoy2020, SinhaBanerjee2021, deCruzetal2021, Xuetal2022, Jesusetal2022, Adiletal2023, Dongetal2023, VanRaamsdonkWaddell2023, Avsajanishvilietal2024, VanRaamsdonkWaddell2024a, Thompson2024} for recent constraints.} where a scalar field $\phi$ represents the dynamical dark energy. The Friedmann equation for this model is given by 
\be
\label{eq:Hzp}
H(z) = H_0\sqrt{\Om\left(1 + z\right)^3 + \Ok\left(1 + z\right)^2 + \Omega_{\phi}(z,\alpha)},
\ee
with the scalar field dynamical dark energy density parameter $\Omega_{\phi}(z,\alpha)$ derived by numerically solving both \eqref{eq:Hzp} and the scalar field equation of motion
\be
\label{em}
\ddot{\phi}+3H\dot{\phi}+V'(\phi)=0.
\ee
Here an overdot represents a derivative with respect to time, the prime represents a derivative with respect to $\phi$, and the scalar field potential energy density $V(\phi)$ is given by
\be
\label{PE}
V(\phi)=\frac{1}{2}\kappa m_p^2\phi^{-\alpha},
\ee
where $m_p$ is the Planck mass, $\alpha$ is a non-negative model parameter ($\alpha=0$ reduces to \lcdm), and $\kappa$ is determined via the shooting method in the cosmic linear anisotropy solving system (\textsc{class}) \cite{class}. For GRB data, we constrain $\Om, \alpha$, and $\Ok$, while for $H(z)$ + BAO data we vary $H_0, \obh\!, \och\!, \alpha$, and $\Ok$. In the spatially flat model, $\Ok=0$.

\section{Data}
\label{sec:data}

We analyze 151 long GRBs (A123 and A28) from Ref.~\cite{WangLiang2024} to test their standardizability via the Amati correlation within six cosmological models. The GRB and comparison $H(z)$ + BAO datasets are summarized below.

\begin{itemize}

\item[]{\it GRB samples.} The A123 and A28 samples, compiled by Ref.\ \cite{WangLiang2024} (tables A1 and A2), consist of 123 and 28 long GRBs from Fermi observations, spanning redshift ranges of $0.117 \leq z \leq 5.6$ and $0.5519 \leq z \leq 8.2$, respectively. Differences in rest-frame peak energy ($E_{\rm p}$, in keV) and bolometric fluence ($S_{\rm bolo}$, in $\text{erg\,cm}^{-2}$) for GRBs common to both the A123 (43 of 123 GRBs) / A28 (6 of 28 GRBs) and A118 (49 of 118 GRBs) \citep{Khadkaetal_2021b} samples are listed in Table \ref{tab:diffs}. Footnotes in the table list the proximate causes of these differences.

\begin{sidewaystable*}
\centering
\resizebox*{0.94\columnwidth}{0.68\columnwidth}{%
\begin{threeparttable}
\caption{Differences in rest-frame peak energy ($E_{\rm p}$) and bolometric fluence ($S_{\rm bolo}$) for GRBs common to both the A123 / A28 (upper / lower part) \citep{WangLiang2024} and A118 \citep{Khadkaetal_2021b} samples.}\label{tab:diffs}
\setlength{\tabcolsep}{10pt}
\begin{tabular}{lcccccccc}
\toprule\toprule
GRB & $\Delta E_{\rm p}$ & $\Delta S_{\rm bolo}$ & GRB & $\Delta E_{\rm p}$ & $\Delta S_{\rm bolo}$ & GRB & $\Delta E_{\rm p}$ & $\Delta S_{\rm bolo}$ \\
\midrule
080810\tnote{a} & $2.64\sigma$ & $3.19\sigma$ & 081121\tnote{a} & $2.30\sigma$ & $0.81\sigma$ & 081222\tnote{a} & $1.07\sigma$ & $1.35\sigma$ \\
090323\tnote{b} & $0.07\sigma$ & $4.47\sigma$ & 090328A\tnote{c} & $0.83\sigma$ & $2.94\sigma$ & 090424\tnote{c} & $1.32\sigma$ & $1.60\sigma$ \\
090516A\tnote{a} & $0.53\sigma$ & $0.27\sigma$ & 090902B\tnote{b} & $16.37\sigma$ & $10.47\sigma$ & 90926A\tnote{b} & $7.38\sigma$ & $3.66\sigma$ \\
091003A\tnote{c} & $2.58\sigma$ & $4.19\sigma$ & 091020\tnote{a} & $1.46\sigma$ & $0.11\sigma$ & 091127\tnote{b} & $2.49\sigma$ & $12.56\sigma$ \\
091208B\tnote{c} & $3.33\sigma$ & $3.21\sigma$ & 100414A\tnote{c} & $4.38\sigma$ & $0.14\sigma$ & 100728A\tnote{c} & $5.90\sigma$ & $8.69\sigma$ \\
100814A\tnote{a} & $0.01\sigma$ & $1.43\sigma$ & 100906A\tnote{a} & $0.72\sigma$ & $0.03\sigma$ & 110213A\tnote{a} & $0.70\sigma$ & $1.48\sigma$ \\
110731A\tnote{a} & $0.85\sigma$ & $2.01\sigma$ & 120119A\tnote{a} & $1.32\sigma$ & $0.32\sigma$ & 120326A\tnote{a} & $0.32\sigma$ & $1.52\sigma$ \\ 
120624B\tnote{b} & $10.41\sigma$ & $12.13\sigma$ & 120811C\tnote{a} & $1.81\sigma$ & $4.63\sigma$ & 120909A\tnote{a} & $3.88\sigma$ & $6.15\sigma$ \\
121128A\tnote{a} & $2.87\sigma$ & $3.03\sigma$ & 130215A\tnote{a} & $0.72\sigma$ & $1.60\sigma$ & 130420A\tnote{a} & $0.68\sigma$ & $4.22\sigma$ \\ 
130427A\tnote{b} & $86.62\sigma$ & $165.04\sigma$ & 130518A\tnote{c} & $4.53\sigma$ & $4.93\sigma$ & 130610A\tnote{a} & $0.07\sigma$ & $6.05\sigma$ \\
131105A\tnote{a} & $1.80\sigma$ & $7.84\sigma$ & 131108A\tnote{b} & $1.34\sigma$ & $1.98\sigma$ & 131231A\tnote{c} & $9.39\sigma$ & $10.62\sigma$ \\
140206A\tnote{a} & $0.14\sigma$ & $1.76\sigma$ & 140213A\tnote{a} & $1.35\sigma$ & $3.95\sigma$ & 141028A\tnote{c} & $4.38\sigma$ & $0.97\sigma$ \\
150314A\tnote{c} & $1.11\sigma$ & $3.70\sigma$ & 150403A\tnote{b} & $6.45\sigma$ & $0.73\sigma$ & 150514A\tnote{c} & $1.15\sigma$ & $2.33\sigma$ \\
160509A\tnote{b} & $28.45\sigma$ & $16.83\sigma$ & 160625B\tnote{b} & $10.23\sigma$ & $9.12\sigma$ & 170214A\tnote{b} & $3.35\sigma$ & $0.89\sigma$ \\
170405A\tnote{c} & $4.03\sigma$ & $2.20\sigma$ & & & & & & \\
\midrule
080916C\tnote{b} & $2.80\sigma$ & $10.33\sigma$ & 090423\tnote{a}  & $0.53\sigma$ & $0.99\sigma$ & 110818A\tnote{a} & $0.95\sigma$ & $3.51\sigma$ \\
111107A\tnote{a} & $1.04\sigma$ & $1.86\sigma$ & 120922A\tnote{a} & $0.54\sigma$ & $2.36\sigma$ & 30612A\tnote{a} & $2.47\sigma$ & $0.14\sigma$ \\
\bottomrule\bottomrule
\end{tabular}
\begin{tablenotes}
\item [a] Differences arise from different broken power laws in the BAND model of prompt emission spectrum $\Phi(E)$ between Ref.~\cite{Wang_2016} and Ref.~\cite{WangLiang2024}.
\item [b] Differences arise from different models of $\Phi(E)$ between Ref.~\cite{Dirirsa2019} (SBPL, BAND+PL, BAND+BB, BAND+CPL+PL, BAND+BB+PL, or SBPL+BB) and Ref.~\cite{WangLiang2024} (BAND).
\item [c] Differences arise from slightly different broken power laws in the BAND model of $\Phi(E)$ between Ref.~\cite{Dirirsa2019} (standard BAND) and Ref.~\cite{WangLiang2024} [missing $(100\,\text{keV})^{\beta-\alpha}$ term for $E > (\alpha-\beta)/(2+\alpha)E_{\rm p,obs}$ case], and different values of $\alpha$ and $\beta$ (photon indices), and $T_{90}$ (duration of 90\% of gamma-ray emission).
\end{tablenotes}
\end{threeparttable}%
}
\end{sidewaystable*}

The Amati correlation \citep{Amatietal2002, Amati2008, Amati2009} is 
\begin{equation}
\label{Amati}
    \log E_{\rm iso} = \beta + \gamma\log E_{\rm p},
\end{equation}
where $\gamma$ and $\beta$ are the slope and intercept parameters, respectively. The rest-frame peak energy is $E_{\rm p} = (1+z)E_{\rm p, obs}$, where $E_{\rm p, obs}$ is the observed peak energy (in KeV). The rest-frame isotropic radiated energy $E_{\rm iso}$ (in erg) is $E_{\rm iso}=4\pi D_L^2S_{\rm bolo}/(1+z)$, where $D_L(z)$ is the luminosity distance
\begin{equation}
  \label{eq:DL}
\resizebox{0.475\textwidth}{!}{%
    $D_L(z) = 
    \begin{cases}
    \frac{c(1+z)}{H_0\sqrt{\Ok}}\sinh\left[\frac{\sqrt{\Ok}H_0}{c}D_C(z)\right] & \text{if}\ \Ok > 0, \\
    \vspace{1mm}
    (1+z)D_C(z) & \text{if}\ \Ok = 0,\\
    \vspace{1mm}
    \frac{c(1+z)}{H_0\sqrt{|\Ok|}}\sin\left[\frac{H_0\sqrt{|\Ok|}}{c}D_C(z)\right] & \text{if}\ \Ok < 0,
    \end{cases}$%
    }
\end{equation}
with $D_C(z)$ being the comoving distance
\begin{equation}
\label{eq:gz}
   D_C(z) = c\int^z_0 \frac{dz'}{H(z')},
\end{equation}
and $c$ being the speed of light.

We test the standardizability of these GRB data by simultaneously fitting the Amati correlation and cosmological parameters within each cosmological model, assessing the consistency of the resulting Amati correlation parameters across models.

\item[]{$H(z)\ +\ BAO\ data$.} For the $H(z)$ + BAO analyses we use 32 $H(z)$ measurements ($0.07 \leq z \leq 1.965$) and 12 BAO measurements ($0.122 \leq z \leq 2.334$) from Ref.~\cite{CaoRatra2023} (tables 1 and 2). In our analyses we account for all known covariancs in these data.

\end{itemize}

\section{Data Analysis Methodology}
\label{sec:analysis}

We perform Bayesian inference analyses using the \textsc{MontePython} Markov chain Monte Carlo (MCMC) code \citep{Audrenetal2013,Brinckmann2019}. Flat (uniform) priors are adopted for all constrained parameters (Table \ref{tab:priors}). Posterior distributions are derived and plotted using the \textsc{GetDist} \textsc{python} package \citep{Lewis_2019}. 

\begin{table*}[htbp]
\centering
\setlength\tabcolsep{24pt}
\begin{threeparttable}
\caption{Flat (uniform) priors of the constrained parameters.}
\label{tab:priors}
\begin{tabular}{lcc}
\toprule\toprule
Parameter & & Prior\\
\midrule
 & Cosmological Parameters & \\
\midrule
$H_0$\,\tnote{a} &  & [None, None]\\
\obhs\,\tnote{b} &  & [0, 1]\\
\ochs\,\tnote{b} &  & [0, 1]\\
\ok &  & [$-2$, 2]\\
$\alpha$ &  & [0, 10]\\
\wx &  & [$-5$, 0.33]\\
\om\,\tnote{c} &  & [0.051314766115, 1]\\
\\
 & Amati Correlation Parameters & \\
$\beta$ &  & [0, 300]\\
$\gamma$ &  & [0, 5]\\
$\sigma_{\mathrm{int}}$ &  & [0, 5]\\
\bottomrule\bottomrule
\end{tabular}
\begin{tablenotes}
\item [a] \hunit. In the GRB cases $H_0$ is set to be 70 \hunit.
\item [b] $H(z)$ + BAO. In the GRB cases $\Omega_{b}$ is set to be $0.05$.
\item [c] GRB cases only, to ensure that $\Omega_{c}$ remains positive.
\end{tablenotes}
\end{threeparttable}%
\end{table*}

The natural logarithm of the GRB likelihood function is given by
\begin{equation}
\label{eq:LF_s1}
    \ln\mathcal{L}_{\rm GRB}= -\frac{1}{2}\Bigg[\chi^2_{\rm GRB}+\sum^{N}_{i=1}\ln\left(2\pi\sigma^2_{\mathrm{tot},i}\right)\Bigg],
\end{equation}
where
\begin{equation}
\label{eq:chi2_s1}
    \chi^2_{\rm GRB} = \sum^{N}_{i=1}\bigg[\frac{(\log E_{\mathrm{iso},i} - \beta - \gamma\log E_{\mathrm{p},i})^2}{\sigma^2_{\mathrm{tot},i}}\bigg]
\end{equation}
with total uncertainty $\sigma_{\mathrm{tot},i}$ given by
\begin{equation}
\label{eq:sigma_s2}
\sigma^2_{\mathrm{tot},i}=\sigma_{\mathrm{int}}^2+\sigma_{\log E_{\mathrm{iso},i}}^2+\gamma^2\sigma_{\log E_{\mathrm{p},i}}^2.
\end{equation}
Here, $\sigma_{\mathrm{int}}$ represents the intrinsic scatter of the GRB data, which also accounts for unknown systematic uncertainties \citep{DAgostini2005}.

The likelihood functions and covariance matrices for the $H(z)$ and BAO data are detailed in Ref.\ \cite{CaoRatra2023}. Following Ref.\ \cite{CaoRatra2023} (see their Sec.~IV for details), we also use the Akaike Information Criterion (AIC), Bayesian Information Criterion (BIC), and Deviance Information Criterion (DIC) to assess the goodness of fit for the cosmological models.

\begin{sidewaystable*}
\centering
\resizebox{1\columnwidth}{!}{%
\begin{threeparttable}
\caption{Unmarginalized best-fitting parameter values for all models from various combinations of data.}\label{tab:BFP}
\begin{tabular}{lccccccccccccccccc}
\toprule\toprule
Model & dataset & $\Omega_{b}h^2$ & $\Omega_{c}h^2$ & $\Omega_{m0}$ & $\Omega_{k0}$ & $w_{\mathrm{X}}$/$\alpha$\tnote{a} & $H_0$\tnote{b} & $\gamma$ & $\beta$ & $\sigma_{\mathrm{int}}$ & $-2\ln\mathcal{L}_{\mathrm{max}}$ & AIC & BIC & DIC & $\Delta \mathrm{AIC}$ & $\Delta \mathrm{BIC}$ & $\Delta \mathrm{DIC}$ \\
\midrule
Flat \lcdm & $H(z)$ + BAO & 0.0254 & 0.1200 & 0.297 & -- & -- & 70.12 & -- & -- & -- & 30.56 & 36.56 & 41.91 & 37.32 & 0.00 & 0.00 & 0.00\\
 & A123\tnote{c} & -- & -- & 1.000 & -- & -- & -- & 1.203 & 49.62 & 0.555 & 212.91 & 218.91 & 227.34 & 222.42 & 0.00 & 0.00 & 0.00\\
 & A28\tnote{c} & -- & -- & 0.997 & -- & -- & -- & 0.735 & 50.32 & 0.648 & 58.02 & 64.02 & 68.02 & 67.35 & 0.00 & 0.00 & 0.00\\
 & A123 + A28\tnote{c} & -- & -- & 1.000 & -- & -- & -- & 1.190 & 49.57 & 0.593 & 285.90 & 291.90 & 300.95 & 295.54 & 0.00 & 0.00 & 0.00\\[6pt]
Nonflat \lcdm & $H(z)$ + BAO & 0.0269 & 0.1128 & 0.289 & 0.041 & -- & 69.61 & -- & -- & -- & 30.34 & 38.34 & 45.48 & 38.80 & 1.78 & 3.56 & 1.48\\
 & A123\tnote{c} & -- & -- & 0.996 & $-1.475$ & -- & -- & 1.076 & 49.97 & 0.542 & 207.46 & 215.46 & 226.70 & 219.73 & $-3.45$ & $-0.64$ & $-2.69$\\
 & A28\tnote{c} & -- & -- & 0.357 & $-1.151$ & -- & -- & 0.034 & 51.68 & 0.513 & 45.68 & 53.68 & 59.01 & 71.50 & $-10.35$ & $-9.01$ & 4.15\\
 & A123 + A28\tnote{c} & -- & -- & 1.000 & $-1.310$ & -- & -- & 1.095 & 49.84 & 0.584 & 281.31 & 289.31 & 301.38 & 293.28 & $-2.59$ & 0.43 & $-2.27$\\[6pt]
Flat XCDM & $H(z)$ + BAO & 0.0320 & 0.0932 & 0.283 & -- & $-0.731$ & 66.69 & -- & -- & -- & 26.57 & 34.57 & 41.71 & 34.52 & $-1.98$ & $-0.20$ & $-2.80$\\
 & A123\tnote{c} & -- & -- & 0.071 & -- & 0.140 & -- & 1.183 & 49.60 & 0.546 & 210.59 & 218.59 & 229.84 & 224.16 & $-0.32$ & 2.49 & 1.74\\
 & A28\tnote{c} & -- & -- & 0.052 & -- & 0.140 & -- & 0.717 & 50.25 & 0.638 & 57.01 & 65.01 & 70.34 & 67.43 & 0.99 & 2.32 & 0.08\\
 & A123 + A28\tnote{c} & -- & -- & 0.052 & -- & 0.139 & -- & 1.156 & 49.58 & 0.594 & 283.69 & 291.69 & 303.76 & 296.85 & $-0.21$ & 2.80 & 1.31\\[6pt]
Nonflat XCDM & $H(z)$ + BAO & 0.0312 & 0.0990 & 0.293 & $-0.085$ & $-0.693$ & 66.84 & -- & -- & -- & 26.00 & 36.00 & 44.92 & 36.17 & $-0.56$ & 3.01 & $-1.15$\\
 & A123\tnote{c} & -- & -- & 0.071 & $-1.993$ & 0.125 & -- & 1.078 & 49.59 & 0.537 & 205.85 & 215.85 & 229.91 & 218.81 & $-3.06$ & 2.56 & $-3.61$\\
 & A28\tnote{c} & -- & -- & 0.467 & $-1.670$ & $-0.844$ & -- & 0.052 & 51.52 & 0.500 & 43.39 & 53.39 & 60.05 & 72.78 & $-10.63$ & $-7.97$ & 5.43\\
 & A123 + A28\tnote{c} & -- & -- & 0.594 & $-1.965$ & 0.123 & -- & 1.058 & 49.56 & 0.584 & 279.82 & 289.82 & 304.91 & 292.01 & $-2.08$ & 3.96 & $-3.53$\\[6pt]
Flat \pcdm & $H(z)$ + BAO & 0.0337 & 0.0864 & 0.271 & -- & 1.169 & 66.78 & -- & -- & -- & 26.50 & 34.50 & 41.64 & 34.01 & $-2.05$ & $-0.27$ & $-3.31$\\
 & A123\tnote{c} & -- & -- & 1.000 & -- & 7.640 & -- & 1.211 & 49.59 & 0.554 & 212.91 & 220.91 & 232.15 & 222.31 & 2.00 & 4.81 & $-0.11$\\
 & A28\tnote{c} & -- & -- & 1.000 & -- & 0.853 & -- & 0.745 & 50.30 & 0.642 & 58.02 & 66.02 & 71.35 & 66.70 & 2.00 & 3.33 & $-0.65$\\
 & A123 + A28\tnote{c} & -- & -- & 0.999 & -- & 1.861 & -- & 1.191 & 49.57 & 0.594 & 285.90 & 293.90 & 305.97 & 295.00 & 2.00 & 5.02 & $-0.54$\\[6pt]
Nonflat \pcdm & $H(z)$ + BAO & 0.0338 & 0.0878 & 0.273 & $-0.077$ & 1.441 & 66.86 & -- & -- & -- & 25.92 & 35.92 & 44.84 & 35.12 & $-0.64$ & 2.93 & $-2.20$\\
 & A123\tnote{c} & -- & -- & 0.992 & $-0.990$ & 9.354 & -- & 1.124 & 49.69 & 0.554 & 208.79 & 218.79 & 232.85 & 221.34 & $-0.12$ & 5.50 & $-1.08$\\
 & A28\tnote{c} & -- & -- & 0.997 & $-0.957$ & 0.160 & -- & 0.646 & 50.54 & 0.604 & 55.39 & 65.39 & 72.05 & 67.83 & 1.36 & 4.03 & 0.49\\
 & A123 + A28\tnote{c} & -- & -- & 0.992 & $-0.983$ & 8.599 & -- & 1.134 & 49.60 & 0.576 & 282.08 & 292.08 & 307.16 & 294.04 & 0.18 & 6.21 & $-1.50$\\
\bottomrule\bottomrule
\end{tabular}
\begin{tablenotes}
\item [a] \wx\ corresponds to flat/nonflat XCDM and $\alpha$ corresponds to flat/nonflat \pcdm.
\item [b] \hunit.
\item [c] $\Omega_b=0.05$ and $H_0=70$ \hunit.
\end{tablenotes}
\end{threeparttable}%
}
\end{sidewaystable*}

\begin{sidewaystable*}
\centering
\resizebox{1\columnwidth}{!}{%
\begin{threeparttable}
\caption{One-dimensional marginalized posterior mean values and uncertainties ($\pm 1\sigma$ error bars or $1\sigma$/$2\sigma$ limits) of the parameters for all models from various combinations of data.}\label{tab:1d_BFP}
\begin{tabular}{lcccccccccc}
\toprule\toprule
Model & dataset & $\Omega_{b}h^2$ & $\Omega_{c}h^2$ & $\Omega_{m0}$ & $\Omega_{k0}$ & $w_{\mathrm{X}}$/$\alpha$\tnote{a} & $H_0$\tnote{b} & $\gamma$ & $\beta$ & $\sigma_{\mathrm{int}}$\\
\midrule
Flat \lcdm & $H(z)$ + BAO & $0.0260\pm0.0040$ & $0.1213^{+0.0091}_{-0.0103}$ & $0.298^{+0.015}_{-0.018}$ & -- & -- & $70.51\pm2.72$ & -- & -- & -- \\
 & A123\tnote{c} & -- & -- & $>0.470$ & -- & -- & -- & $1.222\pm0.132$ & $49.63\pm0.35$ & $0.570^{+0.037}_{-0.046}$ \\%
 & A28\tnote{c} & -- & -- & $>0.213$ & -- & -- & -- & $0.784^{+0.338}_{-0.344}$ & $50.33^{+0.87}_{-0.85}$ & $0.733^{+0.092}_{-0.142}$ \\%
 & A123 + A28\tnote{c} & -- & -- & $>0.471$ & -- & -- & -- & $1.202\pm0.126$ & $49.60\pm0.33$ & $0.607^{+0.037}_{-0.044}$ \\[6pt]%
Nonflat \lcdm & $H(z)$ + BAO & $0.0275^{+0.0047}_{-0.0053}$ & $0.1132\pm0.0183$ & $0.289\pm0.023$ & $0.047^{+0.083}_{-0.091}$ & -- & $69.81\pm2.87$ & -- & -- & -- \\
 & A123\tnote{c} & -- & -- & $>0.503$ & $-0.908^{+0.205}_{-0.599}$ & -- & -- & $1.138\pm0.138$ & $49.86\pm0.36$ & $0.562^{+0.036}_{-0.044}$ \\
 & A28\tnote{c} & -- & -- & $>0.309$ & $-0.778^{+0.238}_{-0.939}$ & -- & -- & $0.648^{+0.284}_{-0.358}$ & $50.57^{+0.87}_{-0.70}$ & $0.685^{+0.088}_{-0.134}$ \\
 & A123 + A28\tnote{c} & -- & -- & $>0.498$ & $-0.784^{+0.215}_{-0.602}$ & -- & -- & $1.136\pm0.131$ & $49.78\pm0.34$ & $0.602^{+0.036}_{-0.043}$ \\[6pt]
Flat XCDM & $H(z)$ + BAO & $0.0308^{+0.0053}_{-0.0046}$ & $0.0980^{+0.0182}_{-0.0161}$ & $0.286\pm0.019$ & -- & $-0.778^{+0.132}_{-0.104}$ & $67.20^{+3.05}_{-3.06}$ & -- & -- & -- \\
 & A123\tnote{c} & -- & -- & $>0.279$ & -- & $<0.090$ & -- & $1.218\pm0.127$ & $49.66\pm0.34$ & $0.569^{+0.037}_{-0.045}$ \\%
 & A28\tnote{c} & -- & -- & $>0.475$\tnote{d} & -- & $<-0.042$ & -- & $0.776^{+0.315}_{-0.350}$ & $50.41\pm0.85$ & $0.727^{+0.090}_{-0.138}$ \\%
 & A123 + A28\tnote{c} & -- & -- & $>0.288$ & -- & $<0.082$ & -- & $1.200\pm0.123$ & $49.63\pm0.33$ & $0.606^{+0.036}_{-0.043}$ \\[6pt]%
Nonflat XCDM & $H(z)$ + BAO & $0.0305^{+0.0055}_{-0.0047}$ & $0.1011\pm0.0196$ & $0.292\pm0.024$ & $-0.059\pm0.106$ & $-0.746^{+0.135}_{-0.090}$ & $67.17^{+2.96}_{-2.97}$ & -- & -- & -- \\
 & A123\tnote{c} & -- & -- & $>0.188$ & $-0.997^{+0.559}_{-0.580}$ & $-1.520^{+1.671}_{-2.007}$ & -- & $1.129\pm0.130$ & $49.84^{+0.37}_{-0.40}$ & $0.559^{+0.035}_{-0.043}$ \\
 & A28\tnote{c} & -- & -- & $>0.223$ & $-0.671^{+0.443}_{-0.700}$ & $-2.253^{+2.310}_{-1.258}$ & -- & $0.651^{+0.278}_{-0.347}$ & $50.63^{+0.89}_{-0.72}$ & $0.679^{+0.085}_{-0.132}$ \\
 & A123 + A28\tnote{c} & -- & -- & $>0.190$ & $-0.908^{+0.603}_{-0.497}$ & $-1.622^{+1.772}_{-1.424}$ & -- & $1.122\pm0.126$ & $49.78^{+0.38}_{-0.37}$ & $0.599^{+0.035}_{-0.042}$ \\[6pt]
Flat \pcdm & $H(z)$ + BAO & $0.0327^{+0.0060}_{-0.0031}$ & $0.0866^{+0.0192}_{-0.0176}$ & $0.272\pm0.022$ & -- & $1.261^{+0.494}_{-0.810}$ & $66.23^{+2.87}_{-2.86}$ & -- & -- & --  \\
 & A123\tnote{c} & -- & -- & $>0.349$ & -- & -- & -- & $1.219\pm0.129$ & $49.61\pm0.34$ & $0.569^{+0.037}_{-0.045}$ \\%
 & A28\tnote{c} & -- & -- & $>0.491$\tnote{d} & -- & -- & -- & $0.782^{+0.335}_{-0.340}$ & $50.29\pm0.85$ & $0.730^{+0.091}_{-0.141}$ \\%
 & A123 + A28\tnote{c} & -- & -- & $>0.355$ & -- & -- & -- & $1.201\pm0.125$ & $49.58\pm0.32$ & $0.606^{+0.037}_{-0.044}$ \\[6pt]%
Nonflat \pcdm & $H(z)$ + BAO & $0.0324^{+0.0062}_{-0.0031}$ & $0.0900\pm0.0200$ & $0.277\pm0.025$ & $-0.072^{+0.093}_{-0.107}$ & $1.435^{+0.579}_{-0.788}$ & $66.50\pm2.88$ & -- & -- & -- \\
 & A123\tnote{c} & -- & -- & $>0.467$ & $-0.452^{+0.214}_{-0.361}$ & $>3.904$\tnote{d} & -- & $1.189\pm0.128$ & $49.64\pm0.33$ & $0.564^{+0.036}_{-0.044}$ \\
 & A28\tnote{c} & -- & -- & $>0.239$ & $-0.258^{+0.316}_{-0.395}$ & -- & -- & $0.765^{+0.312}_{-0.348}$ & $50.28\pm0.83$ & $0.717^{+0.089}_{-0.137}$ \\%
 & A123 + A28\tnote{c} & -- & -- & $>0.463$ & $-0.445^{+0.219}_{-0.356}$ & $>3.820$\tnote{d} & -- & $1.172\pm0.124$ & $49.61\pm0.32$ & $0.602^{+0.036}_{-0.043}$ \\
\bottomrule\bottomrule
\end{tabular}
\begin{tablenotes}
\item [a] \wx\ corresponds to flat/nonflat XCDM and $\alpha$ corresponds to flat/nonflat \pcdm.
\item [b] \hunit.
\item [c] $\Omega_b=0.05$ and $H_0=70$ \hunit.
\item [d] This is the 1$\sigma$ limit. The 2$\sigma$ limit is set by the prior and not shown here.
\end{tablenotes}
\end{threeparttable}%
}
\end{sidewaystable*}

\section{Results}
\label{sec:results}

\begin{figure*}[htbp]
\centering
 \subfloat[Flat \lcdm]{%
    \includegraphics[width=0.45\textwidth,height=0.35\textwidth]{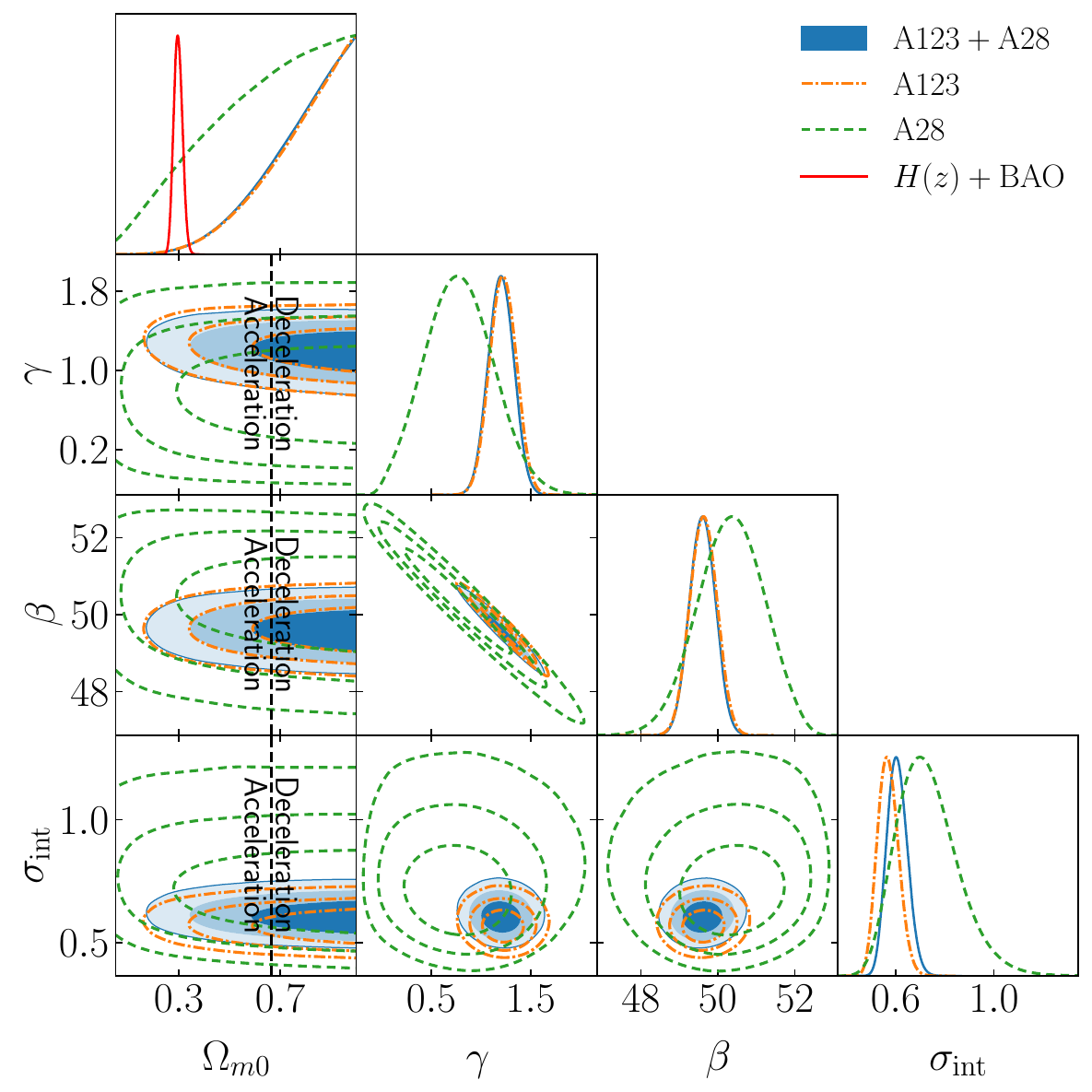}}
 \subfloat[Nonflat \lcdm]{%
    \includegraphics[width=0.45\textwidth,height=0.35\textwidth]{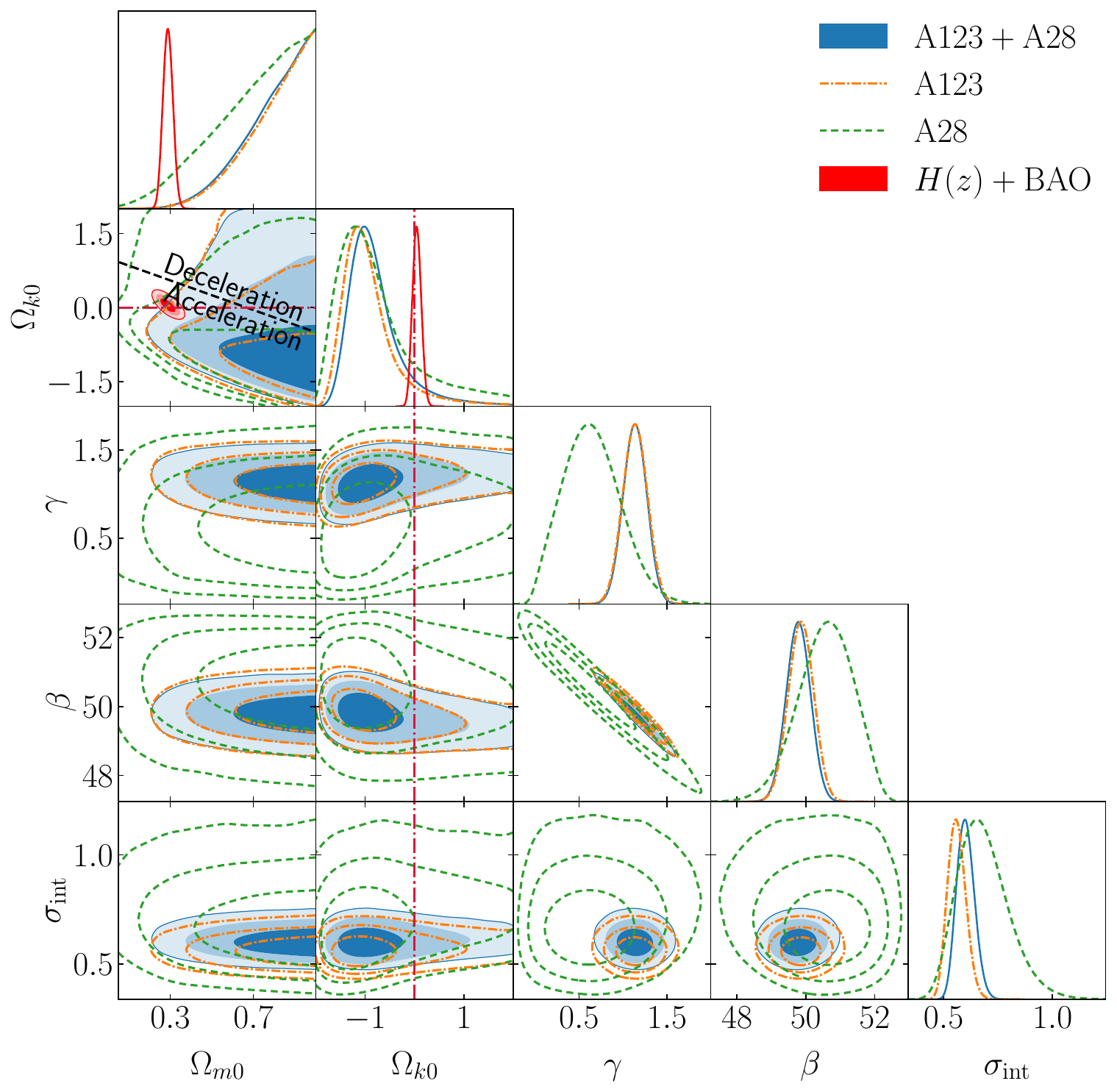}}\\
 \subfloat[Flat XCDM]{%
    \includegraphics[width=0.45\textwidth,height=0.35\textwidth]{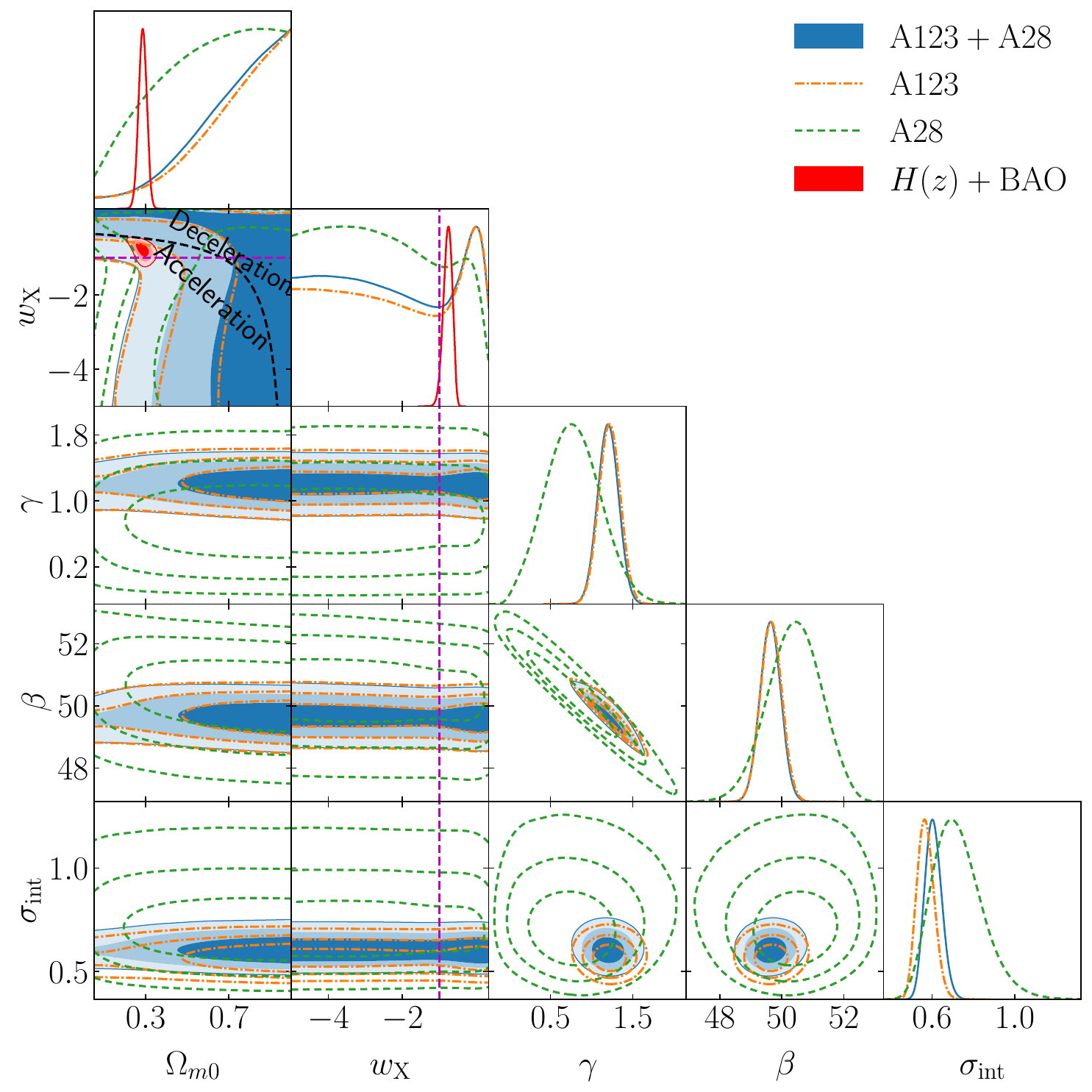}}
 \subfloat[Nonflat XCDM]{%
    \includegraphics[width=0.45\textwidth,height=0.35\textwidth]{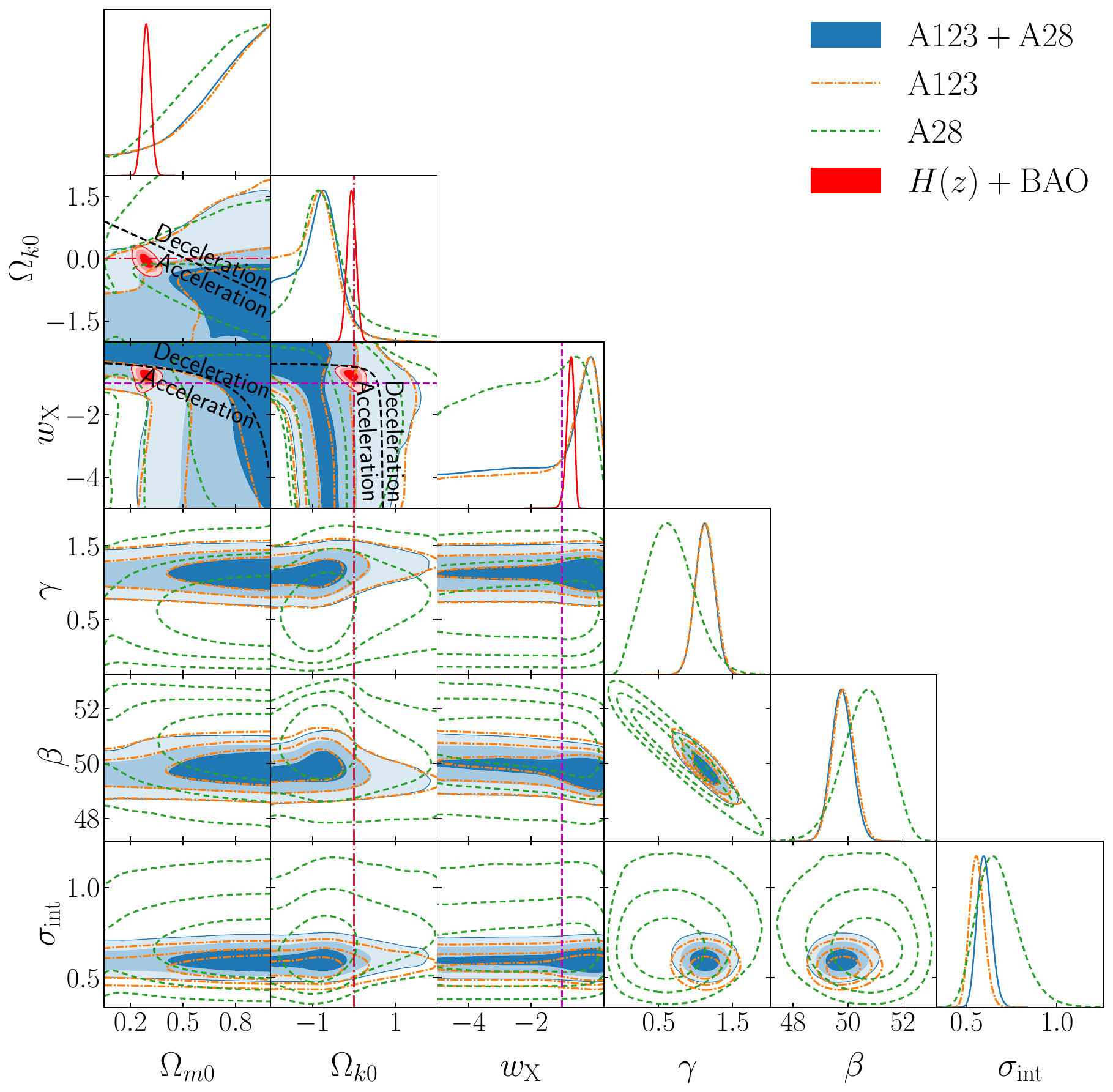}}\\
 \subfloat[Flat \pcdm]{%
    \includegraphics[width=0.45\textwidth,height=0.35\textwidth]{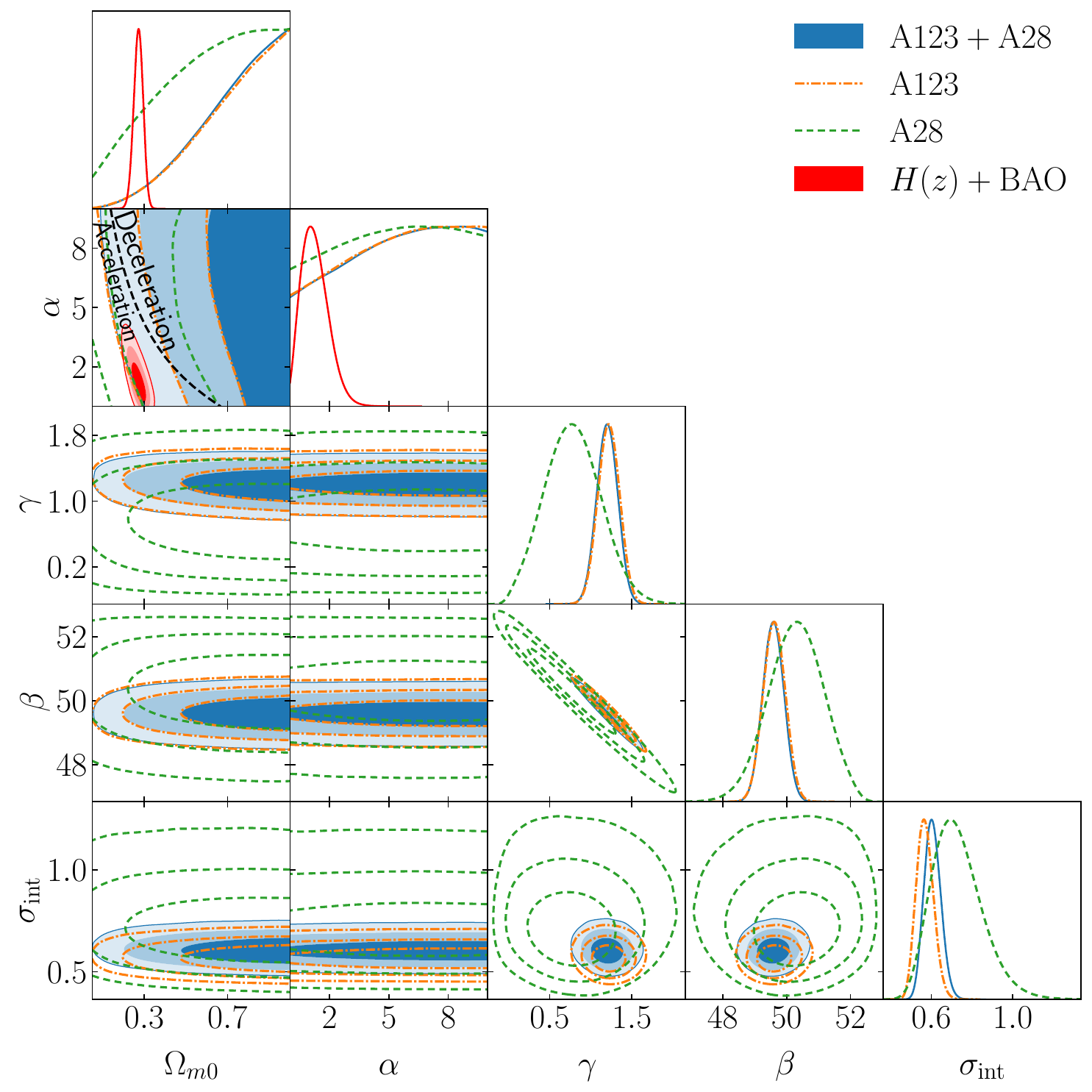}}
 \subfloat[Nonflat \pcdm]{%
    \includegraphics[width=0.45\textwidth,height=0.35\textwidth]{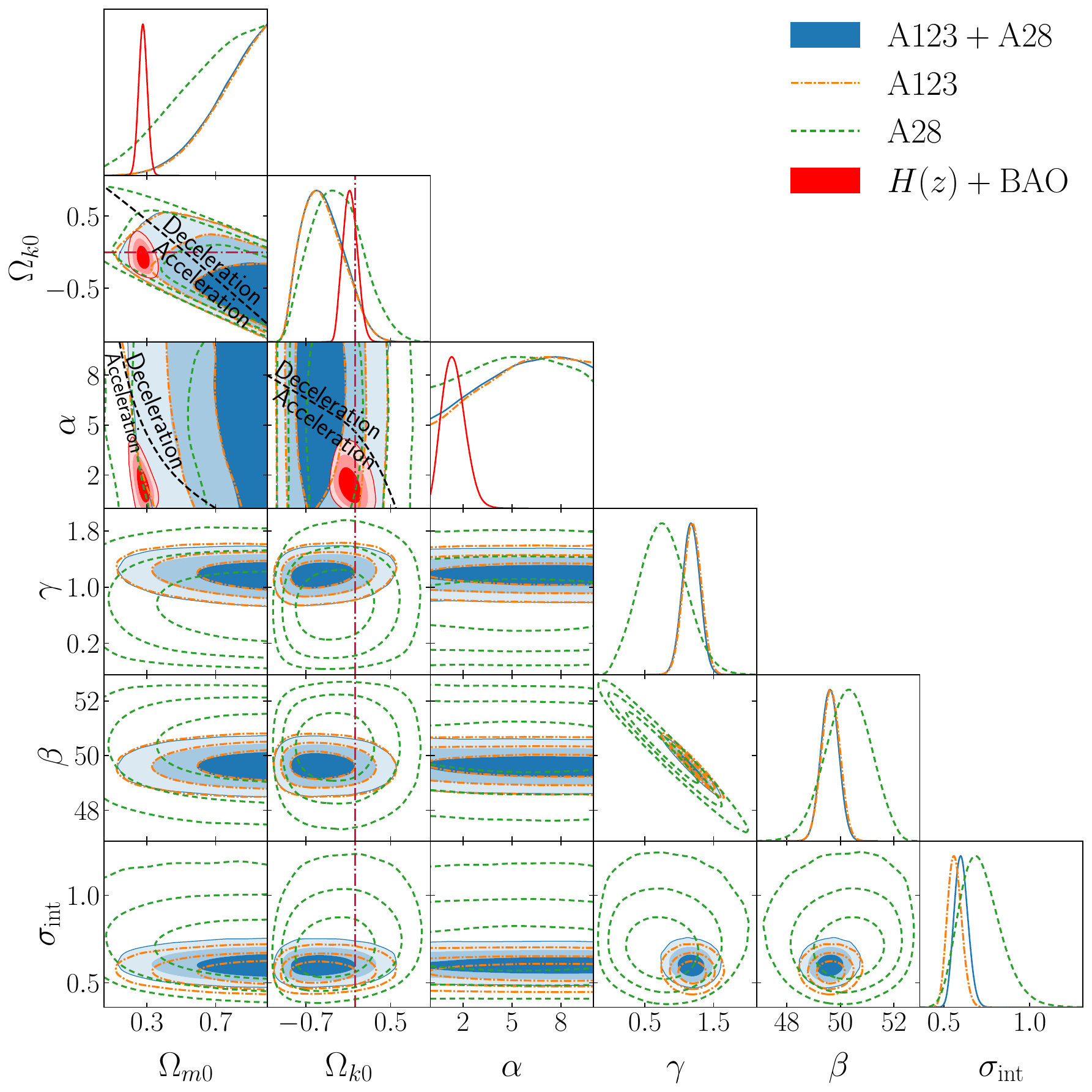}}\\
\caption{One-dimensional likelihoods and 1$\sigma$, 2$\sigma$, and 3$\sigma$ two-dimensional likelihood confidence contours from GRB A123 + A28 (solid blue), A123 (dash-dotted orange), A28 (dashed green), and $H(z)$ + BAO (solid red) data for six different models, with \lcdm, XCDM, and \pcdm\ in the top, middle, and bottom rows, and flat (nonflat) models in the left (right) column. The black dashed zero-acceleration lines, computed for the third cosmological parameter set to the $H(z)$ + BAO data best-fitting values listed in Table \ref{tab:BFP} in panels (d) and (f), divide the parameter space into regions associated with currently-accelerating (below or below left) and currently-decelerating (above or above right) cosmological expansion. The crimson dash-dot lines represent flat hypersurfaces, with closed spatial hypersurfaces either below or to the left. The magenta lines represent $w_{\rm X}=-1$, i.e.\ flat or nonflat \lcdm\ models. The $\alpha = 0$ axes correspond to flat and nonflat \lcdm\ models in panels (e) and (f), respectively.}
\label{fig1}
\vspace{-50pt}
\end{figure*}

\begin{figure*}
\centering
 \subfloat[Flat \lcdm]{%
    \includegraphics[width=0.4\textwidth,height=0.35\textwidth]{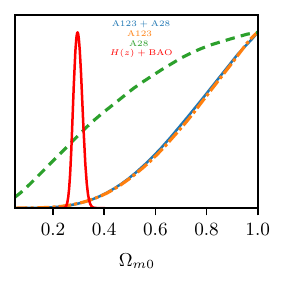}}
 \subfloat[Nonflat \lcdm]{%
    \includegraphics[width=0.4\textwidth,height=0.35\textwidth]{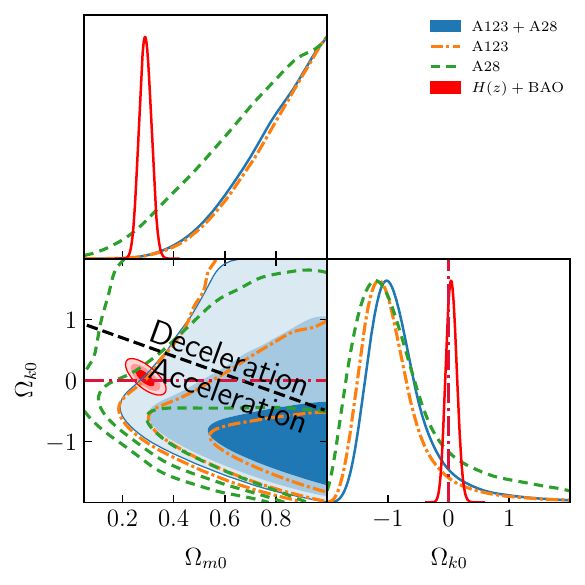}}\\
 \subfloat[Flat XCDM]{%
    \includegraphics[width=0.4\textwidth,height=0.35\textwidth]{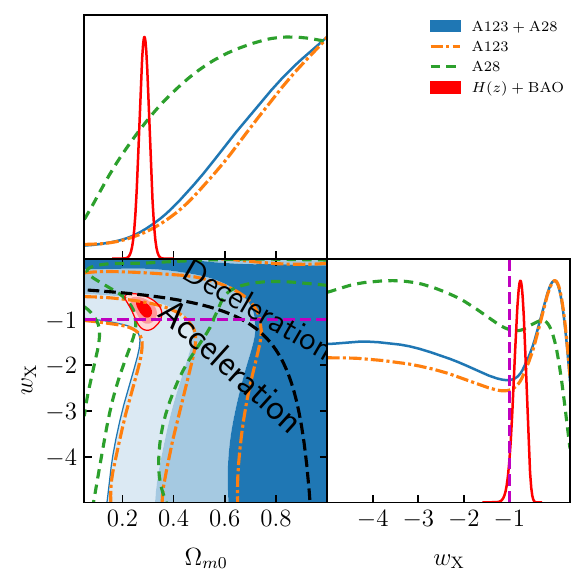}}
 \subfloat[Nonflat XCDM]{%
    \includegraphics[width=0.4\textwidth,height=0.35\textwidth]{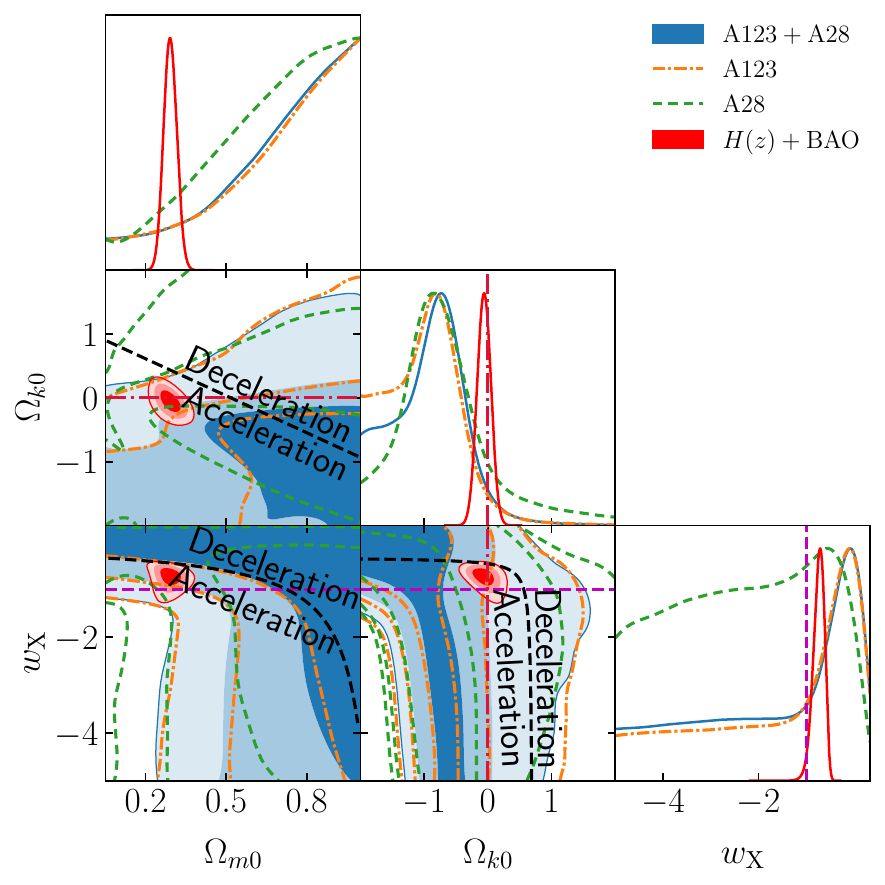}}\\
 \subfloat[Flat \pcdm]{%
    \includegraphics[width=0.4\textwidth,height=0.35\textwidth]{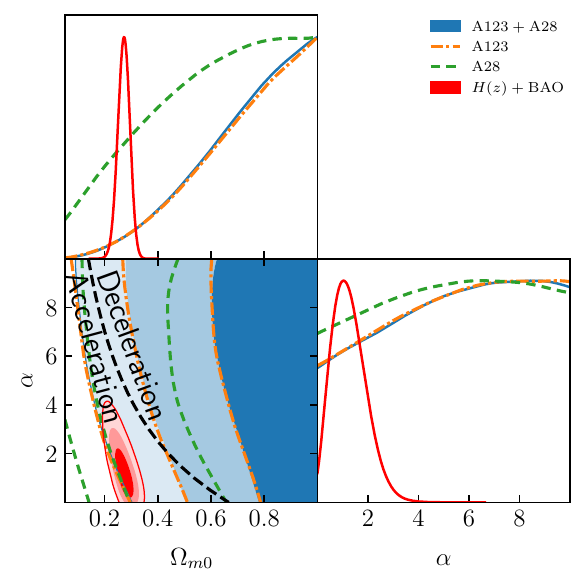}}
 \subfloat[Nonflat \pcdm]{%
    \includegraphics[width=0.4\textwidth,height=0.35\textwidth]{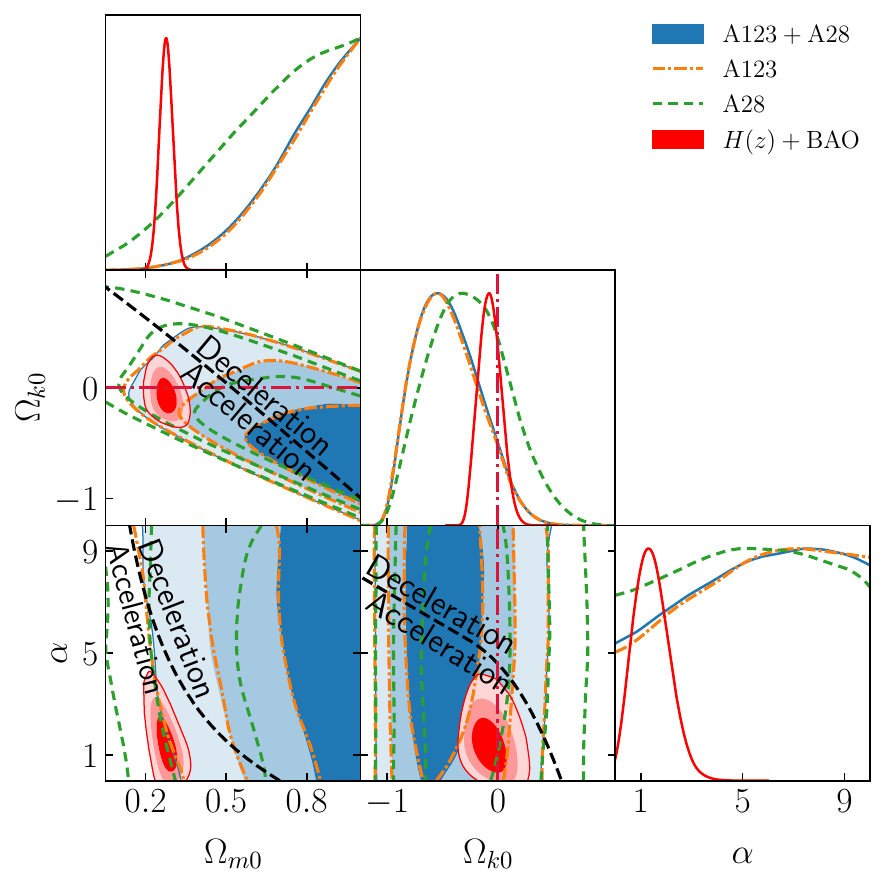}}\\
\caption{Same as Fig.\ \ref{fig1}, but for cosmological parameters only.}
\label{fig2}
\end{figure*}

We present the best-fitting unmarginalized parameter values, maximum likelihood $\mathcal{L}_{\rm max}$ values, AIC, BIC, DIC, $\Delta \mathrm{AIC}$, $\Delta \mathrm{BIC}$, and $\Delta \mathrm{DIC}$ values (where $\Delta$IC values are computed relative to the corresponding flat \lcdm\ model IC values) for all models and datasets in Table \ref{tab:BFP}. The corresponding one-dimensional marginalized posterior distributions values, including $\pm 1\sigma$ uncertainties or 1$\sigma$/2$\sigma$ limits, are provided in Table \ref{tab:1d_BFP}. Figures \ref{fig1} and \ref{fig2} show the posterior distributions and contours for all parameters and for the cosmological parameters only, respectively, for the six cosmological models. Results from the A28, A123, A123 + A28, and $H(z)$ + BAO data are shown in dashed/unfilled green, dash-dotted/unfilled orange, solid/filled blue, and solid/filled red lines/regions, respectively.

To assess the standardizability of the A123 and A28 datasets, we first examine the maximum variations in their Amati correlation and intrinsic scatter parameters across the six studied cosmological models (Table~\ref{tab:diff1}). Because these variations are all within $0.5\sigma$ we conclude that each dataset can be independently standardized using its respective Amati correlation. To establish whether or not the A123 and A28 GRBs follow the same Amati correlation, and so whether or not we can jointly analyze these datasets, we compare the Amati correlation and cosmological parameter constraints for A123 and A28 within each cosmological model, finding consistency within $1.6\sigma$ (Table~\ref{tab:diff2}) and $2\sigma$ (Table~\ref{tab:1d_BFP}), respectively. This consistency justifies combining these datasets into a joint A123 + A28 dataset. We then examine the maximum variations for this combined dataset in Table~\ref{tab:diff1}, finding that it also exhibits small variations (within $0.45\sigma$), confirming its standardizability using the same Amati correlation.

The slope parameter $\gamma$ for A123 ranges from $1.129\pm0.130$ in nonflat XCDM to $1.222\pm0.132$ in flat \lcdm, while for A28 it ranges from $0.648^{+0.284}_{-0.358}$ in nonflat \lcdm\ to $0.784^{+0.338}_{-0.344}$ in flat \lcdm. For the joint A123 + A28 dataset $\gamma$ ranges from $1.122\pm0.126$ in nonflat XCDM to $1.202\pm0.126$ in flat \lcdm.

The intercept parameter $\beta$ for A123 ranges from $49.61\pm0.34$ in nonflat XCDM to $49.86\pm0.36$ in nonflat \lcdm, while for A28 it ranges from $50.28\pm0.83$ in nonflat \pcdm\ to $50.63^{+0.89}_{-0.72}$ in nonflat XCDM. For the joint A123 + A28 dataset $\beta$ ranges from $49.58\pm0.32$ in flat \pcdm\ to $49.78^{+0.38}_{-0.37}$ in nonflat XCDM.

The intrinsic scatter parameter $\sigma_{\rm int}$ for A123 ranges from $0.559^{+0.035}_{-0.043}$ in nonflat XCDM to $0.570^{+0.037}_{-0.046}$ in flat \lcdm, while for A28 it ranges from $0.679^{+0.085}_{-0.132}$ in nonflat XCDM to $0.733^{+0.092}_{-0.142}$ in flat \lcdm, with higher scatter and scatter uncertainties. For the joint A123 + A28 dataset $\sigma_{\rm int}$ ranges from $0.599^{+0.035}_{-0.042}$ in nonflat XCDM to $0.607^{+0.037}_{-0.044}$ in flat \lcdm.

Among the six cosmological models analyzed, the nonflat \lcdm\ model and the $\Ok-\Om$ and $\wX-\Ok$ planes of the nonflat XCDM parametrization show stronger preference for currently accelerating cosmological expansion with the A123 and A123 + A28 datasets. The A28 dataset also more favors accelerating expansion in the nonflat \lcdm, flat XCDM, and nonflat XCDM cases. In the remaining cases, decelerating expansion is preferred, although accelerating expansion remains within the $2\sigma$ confidence region.

We next summarize the cosmological parameter constraints. Compared to the better-established $H(z)$ + BAO data, the A123, A28, and A123 + A28 datasets provide only weak constraints on cosmological parameters. The A28 constraints are consistent with those from the $H(z)$ + BAO data, whereas the A123 and A123 + A28 constraints are mostly not.

A28 data provide only lower limits on \om, ranging from a low of 0.475 ($1\sigma$, flat XCDM) to a high of 0.309 ($2\sigma$, nonflat \lcdm). The \ok\ constraints are $-0.778^{+0.238}_{-0.939}$ (nonflat \lcdm), $-0.671^{+0.443}_{-0.700}$ (nonflat XCDM), and $-0.258^{+0.316}_{-0.395}$ (nonflat \pcdm), all favoring closed hypersurfaces, with spatial flatness consistent within $2\sigma$ ($0.82\sigma$ for nonflat \pcdm). These data do not provide $\alpha$ constraints, and the \wx\ constraints are weak, but all consistent with flat \lcdm\ within $2\sigma$. Although the A28 constraints are consistent with the $H(z)$ + BAO constraints the limited size of the A28 sample (only 28 data points) makes them a less compelling addition for joint analyses, especially compared to A118 (intrinsic scatter $\sim0.7$ of A28 here vs.\ $\sim0.4$ of A118).

For the A123 and A123 + A28 datasets, the $2\sigma$ lower limits on \om\ range from a low of 0.188 (nonflat XCDM) to a high of 0.503 (nonflat \lcdm) and from a low of 0.190 (nonflat XCDM) to a high of 0.498 (nonflat \lcdm), respectively. These limits are consistent within $2\sigma$ with those from $H(z)$ + BAO data only in the flat and nonflat XCDM parametrizations. The \ok\ constraints are $-0.908^{+0.205}_{-0.599}$ and $-0.784^{+0.215}_{-0.602}$ in nonflat \lcdm, $-0.997^{+0.559}_{-0.580}$ and $-0.908^{+0.603}_{-0.497}$ in nonflat XCDM, and $-0.452^{+0.214}_{-0.361}$ and $-0.445^{+0.219}_{-0.356}$ in nonflat \pcdm\ for A123 and A123 + A28 data, respectively; all favoring closed hypersurfaces, with spatial flatness within $2\sigma$ except for nonflat XCDM. Both datasets do not provide $\alpha$ constraints for flat \pcdm, and the resulting \wx\ constraints are weak with the flat \lcdm\ ($\wX = -1$) value within $2\sigma$. Due to the inconsistency in the \om\ constraints, neither the A123 nor the A123 + A28 datasets are suitable for joint analyses with $H(z)$ + BAO data, and these GRB datasets should also not be jointly analyzed (or calibrated) with SNIa data, \cite{WangLiang2024}.

Based on the more reliable DIC, the A123 and A123 + A28 datasets favor nonflat XCDM the most, with weak or positive evidence against the remaining models and parametrizations. The A28 data, however, favor flat \pcdm\ the most, showing mildly strong evidence against nonflat XCDM ($\Delta\text{DIC}=6.08$) and weak or positive evidence against other models.

\begin{table*}
\centering
\setlength\tabcolsep{36pt}
\begin{threeparttable}
\caption{The largest differences between $\gamma$, $\beta$, and $\sigma_{\mathrm{int}}$ in the six considered cosmological models (flat and non-flat \lcdm, XCDM, and \pcdm) from various combinations of GRB data, with $1\sigma$ being the quadrature sum of the two corresponding $1\sigma$ uncertainties.}\label{tab:diff1}
\begin{tabular}{lccc}
\toprule
 Data set & $\Delta\gamma$ & $\Delta\beta$ & $\Delta\sigma_{\mathrm{int}}$ \\
\midrule
A123 & $0.50\sigma$ & $0.50\sigma$ & $0.19\sigma$ \\
A28 & $0.30\sigma$ & $0.32\sigma$ & $0.33\sigma$ \\
A123 + A28 & $0.45\sigma$ & $0.41\sigma$ & $0.14\sigma$\\
\bottomrule
\end{tabular}
\end{threeparttable}%
\end{table*}

\begin{table*}
\centering
\setlength\tabcolsep{35pt}
\begin{threeparttable}
\caption{The differences between A123 and A28 for a given cosmological model with $1\sigma$ being the quadrature sum of the two corresponding $1\sigma$ uncertainties.}\label{tab:diff2}
\begin{tabular}{lccc}
\toprule
 Model & $\Delta\gamma$ & $\Delta\beta$ & $\Delta\sigma_{\mathrm{int}}$ \\
\midrule
Flat \lcdm & $1.21\sigma$ & $0.76\sigma$ & $1.11\sigma$ \\
Nonflat \lcdm & $1.55\sigma$ & $0.90\sigma$ & $0.89\sigma$ \\
Flat XCDM & $1.30\sigma$ & $0.82\sigma$ & $1.11\sigma$ \\
Nonflat XCDM & $1.56\sigma$ & $0.98\sigma$ & $0.88\sigma$ \\
Flat \pcdm & $1.22\sigma$ & $0.74\sigma$ & $1.10\sigma$ \\
Nonflat \pcdm & $1.26\sigma$ & $0.72\sigma$ & $1.08\sigma$ \\
\bottomrule
\end{tabular}
\end{threeparttable}%
\end{table*}

\vspace{1cm}
\section{Conclusion}
\label{sec:conclusion}

We use 151 Fermi long GRBs (A123 and A28), compiled by Ref.~\cite{WangLiang2024}, to simultaneously constrain the Amati correlation and cosmological parameters in six spatially flat and nonflat relativistic dark energy cosmological models or parametrizations. Because both the Amati correlation and cosmological parameter constraints from the A123 and A28 datasets are mutually consistent, and because the Amati correlation parameter constraints are independent of the cosmological models, we also perform a joint analysis of the A123 + A28 data. 

We find that the A123, A28, and A123 + A28 datasets are standardizable using the same Amati correlation. However, we also find that the A123 and A123 + A28 constraints on \om\ exhibit a $>2\sigma$ tension with those from better-established $H(z)$ + BAO data for the four flat and nonflat \lcdm\ and \pcdm\ models. Therefore, these A123 and A123 + A28 GRB data cannot be used to constrain cosmological parameters in combination with $H(z)$ + BAO data, nor can they be used jointly with SNIa data, \cite{WangLiang2024}.

Although the A28 constraints are consistent with the $H(z)$ + BAO constraints, their limited sample size (28 GRBs) and significantly higher intrinsic scatter ($\sim0.7$) compared to A118 ($\sim0.4$) \citep{Khadkaetal_2021b} render them significantly less statistically powerful than the A118 dataset. Therefore, the A118 GRB dataset \citep{Khadkaetal_2021b, LuongoMuccino2021, CaoKhadkaRatra2021, CaoDainottiRatra2022, Liuetal2022} is still the most suitable dataset for cosmological purposes.

Finally, given that about a third (49 of 151) of A123 + A28 GRBs are in common with about 40\% (49 of 118) of A118 GRBs, and that 28 of these 49 GBS have at least one of $\Delta E_{\rm p}$ or $\Delta S_{\rm bolo}$ $> 3\sigma$, it would be interesting to determine how large a contribution this makes to the difference between the A123 + A28 and the A118 constraints.

\acknowledgments

The computations for this project were performed on the Beocat Research Cluster at Kansas State University, which is funded in part by NSF grants CNS-1006860, EPS-1006860, EPS-0919443, ACI-1440548, CHE-1726332, and NIH P20GM113109.



\bibliographystyle{JHEP}
\bibliography{biblio.bib}


\end{document}